\documentclass[aps,twocolumn,floatfix,superscriptaddress,prl]{revtex4-1}
\usepackage{amssymb}
\usepackage{graphicx}
\usepackage{dcolumn}
\usepackage{amsmath}
\usepackage{bm}
\usepackage{yhmath}
\usepackage{textcomp}
\usepackage{epstopdf}
\usepackage{color}
\usepackage{ulem}
\usepackage[toc,page,title,titletoc,header]{appendix}
\usepackage[colorlinks,
            linkcolor=blue,
            anchorcolor=blue,
            citecolor=blue
            ]{hyperref}

\begin{document}

\preprint{APS/123-QED}

\title{Self-bound crystals of antiparallel dipolar mixtures}
\author{Maria Arazo}
\affiliation{Departament de F\'isica Qu\`antica i Astrof\'isica, 
Universitat de Barcelona, Mart\'i i Franqu\`es 1, 08028 Barcelona, Spain}
\affiliation{Institut de Ci\`encies del Cosmos, 
Universitat de Barcelona, Mart\'i i Franqu\`es 1, 08028 Barcelona, Spain}
\author{Albert Gallem\'i}
\affiliation{Institut f\"ur Theoretische Physik, 
Leibniz Universit\"at Hannover, 30167 Hannover, Germany}
\author{Montserrat Guilleumas}
\affiliation{Departament de F\'isica Qu\`antica i Astrof\'isica, 
Universitat de Barcelona, Mart\'i i Franqu\`es 1, 08028 Barcelona, Spain}
\affiliation{Institut de Ci\`encies del Cosmos, 
Universitat de Barcelona, Mart\'i i Franqu\`es 1, 08028 Barcelona, Spain}
\author{Ricardo Mayol}
\affiliation{Departament de F\'isica Qu\`antica i Astrof\'isica, 
Universitat de Barcelona, Mart\'i i Franqu\`es 1, 08028 Barcelona, Spain}
\affiliation{Institut de Ci\`encies del Cosmos, 
Universitat de Barcelona, Mart\'i i Franqu\`es 1, 08028 Barcelona, Spain}
\author{Luis Santos}
\affiliation{Institut f\"ur Theoretische Physik, 
Leibniz Universit\"at Hannover, 30167 Hannover, Germany}%

\date{\today}


\begin{abstract}
Recent experiments have created supersolids of dipolar quantum droplets. 
The resulting crystals lack, however, a genuine cohesive energy and are 
maintained by the presence of an external confinement, bearing a resemblance 
to the case of ion Coulomb crystals. We show that a mixture of two 
antiparallel dipolar condensates allows for the creation of potentially 
large, self-bound crystals which, resembling ionic crystals in solid-state physics, 
are maintained by the mutual dipolar attraction between 
the components, with no need of transversal confinement. This opens 
intriguing novel possibilities, including three-dimensionally self-bound 
droplet-ring structures, stripe/labyrinthic patterns, and self-bound 
crystals of droplets surrounded by an 
interstitial superfluid, resembling the case of superfluid Helium in 
porous media.
\end{abstract}

\maketitle



Solid-state crystals are held together by the interplay between different forms of 
attractive and repulsive interactions between their constituents~\cite{Kittel-Book}. 
This interplay results in a finite cohesive or binding energy, defined as the energy 
that must be added to the crystal to separate its components infinitely apart. In the 
presence of an external confinement, crystals may form even in the absence of genuine 
cohesion. A prominent example is provided by trapped ions, which form crystals due to 
the combination of repulsive Coulomb interactions and external confinement~\cite{Thompson2015}. 
There is, however, no cohesive energy, and ion Coulomb crystals unravel in the absence 
of the trap. 

This feature is shared by recently created crystals of quantum droplets 
in dipolar Bose-Einstein condensates~\cite{Boettcher2021,Chomaz2023}. Self-bound 
droplets, elongated along the dipole direction, result from the 
quasi-cancellation of contact and dipolar interactions, and the stabilizing 
effect of quantum fluctuations~\cite{Petrov2015, Kadau2016, Chomaz2016, Schmitt2016}. 
In the presence of confinement along the dipole direction, energy is 
minimized by the creation of multiple droplets, which, in the presence 
of an external confinement perpendicular to the dipole orientation (transversal 
trap), arrange forming a crystal~\cite{Wenzel2017} that may present supersolid 
properties~\cite{Tanzi2019, Boettcher2019, Chomaz2019, Natale2019, Tanzi2019b, Guo2019, Norcia2021, Bland2022a}. 
Similar to the case of ions in Coulomb crystals, droplets repel each 
other. There is hence no genuine cohesive energy of the droplet crystal~(or 
of any other possible density 
pattern~\cite{Zhang2019, Zhang2021,Hertkorn2021, Poli2021}). The transversal 
trap is crucial to keep it bound.

\begin{figure}[t]
\centering
\includegraphics[width=\linewidth]{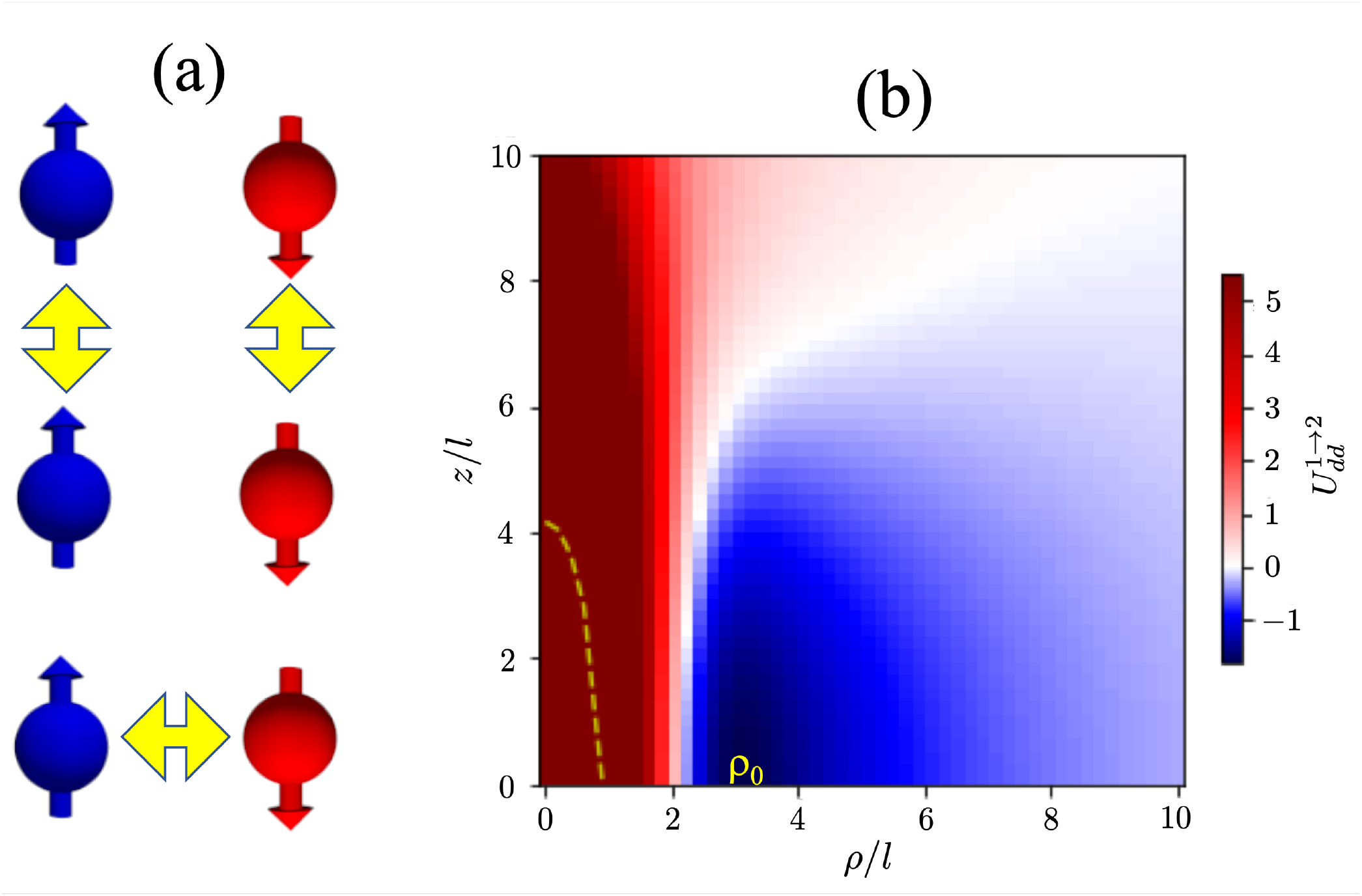}
\caption{(a) In an ADM, intra-component interactions are attractive~(yellow arrows) when 
the particles are head-to-tail, and repulsive when they are side-by-side, whereas the 
opposite is true for the inter-component ones.
(b) Dipolar interaction 
$V^{1\to 2}_{\rm dd}({\bf r})=\frac{2\mu_0\mu_1^2}{3\pi l_z^2} U^{1\to 2}_{\rm dd}$ 
that component 1 exerts on component 2, as a function of $z$ and 
$\rho=\sqrt{x^2+y^2}$. For simplicity, we have assumed a Gaussian droplet 
$e^{-z^2/l_z^2}e^{-\rho^2/2l^2}$. The dashed line indicates the 
half-width-at-half-maximum of the droplet. The inter-component dipolar 
interaction results in an energy minimum on the $xy$ plane at a given 
radius $\rho_0$ well outside the droplet. 
}
\label{fig:1}
\end{figure}

Recent experiments have created a mixture of two dipolar 
components~\cite{Trautmann2018,Durastante2020,Politi2022}. 
These mixtures are expected to present rich physics due to 
the competition between intra- and inter-component contact 
and dipolar interactions, including immiscible droplets~\cite{Smith2021, Bisset2021}, 
doping-induced droplet nucleation~\cite{Politi2022, Scheiermann2023}, 
two-fluid supersolidity~\cite{Scheiermann2023}, and the formation 
of alternating-domain supersolids~\cite{Li2022,Bland2022b,Kirkby2023}. 
Interestingly, the dipoles of the two components may be antiparallel, 
and hence the inter- and intra-component interactions may have opposite 
sign~\cite{Bland2022b}~(Fig.~\ref{fig:1}(a)).



In this letter, we investigate crystal formation in an antiparallel dipolar 
mixture~(ADM). As for a parallel one~\cite{Smith2021,Bisset2021}, in the absence of any 
confinement, an ADM may form an immiscible three-dimensionally self-bound 
mixture, although with a markedly different topology in which one of the 
components may eventually form a ring around a droplet of the other. The 
presence of confinement along the dipole direction results in crystal formation. 
In stark contrast to both single-component dipolar condensates and parallel binary 
mixtures, in an ADM the crystal has a genuine cohesive energy, remaining self-bound 
in the absence of a transversal trap due to the mutual attraction between the components. 
This resembles the case of ionic crystals in solid-state physics, where ions of opposite 
charge arrange in an intertwined crystalline structure bound by their mutual electrostatic 
interaction~\cite{Kittel-Book}. However, the resulting self-bound ADM is not given by two 
intertwined droplet arrays. Symmetric ADMs with similar intra-component interaction 
strengths form self-bound stripe/labyrinthic density patterns. In contrast, in sufficiently 
asymmetric ADMs, one of the components forms an incoherent droplet crystal with an 
approximate triangular structure, whereas the second one remains superfluid and fills 
the lattice interstitials, resembling to some extent superfluid Helium in porous 
media~\cite{Reppy1992}. 


\paragraph{Model.--} We consider a bosonic ADM, with dipoles oriented, 
respectively, along and antiparallel to the $z$ axis. The components 
may belong to the same species or to two different ones. In order to 
illustrate the possible physics, we consider a dysprosium mixture, with 
magnetic dipoles $\mu_1=10\,\mu_B$ and $\mu_2=-10\,\mu_B$, with $\mu_B$ 
the Bohr magneton. Short-range interactions are characterized by the 
intra- and inter-component scattering lengths: $a_{11}$,  $a_{22}$, and 
$a_{12}$. The physics of the mixture is well described by the extended 
Gross-Pitaevskii equation~\cite{Smith2021,Bisset2021}: 
\begin{eqnarray}
i\hbar \dot\Psi_\sigma({\bf r}, t) &=& \left [\frac{-\hbar^2\nabla^2}{2m}+
V_{\rm trap}({\bf r})+\sum_{\sigma'} 
g_{\sigma\sigma'}|\Psi_{\sigma'}({\bf r}, t)|^2 \right\delimiter 0 \nonumber \\
&+&\sum_{\sigma'} \int d^3r' V^{\sigma\sigma'}_{\rm dd}({\bf r}-{\bf r}')
|\Psi_{\sigma'}({\bf r}', t)|^2 \nonumber\\
&+&  \mu_{{\rm LHY},\sigma}[n_{1,2}({\bf r},t)] \Big ]\Psi_\sigma({\bf r}, t)\,,
\label{eq:eGPE}
\end{eqnarray}
where $\Psi_\sigma({\bf r}, t)$ is the condensate wavefunction 
of component $\sigma=1,2$,  $n_\sigma=|\Psi_\sigma|^2$, and 
$g_{\sigma\sigma'}=4\pi\hbar^2a_{\sigma\sigma'}/m$, with $m$ 
the mass of the bosons. The atoms are confined, if at all, only 
along the $z$ axis by a potential $V_{\rm trap}({\bf r})=\frac{1}{2}m\omega_z^2 z^2$. 
The dipole-dipole interaction is given by the potential 
$V^{\sigma\sigma'}_{\rm dd}({\bf r})=\frac{\mu_0\mu_\sigma \mu_{\sigma'}}{4\pi r^3} \left ( 1 - 3\cos^2\theta \right )$, with $\theta$ the angle sustained by the 
$z$ axis and ${\bf r}$. The effect of quantum fluctuations is 
provided by the Lee-Huang-Yang~(LHY) term 
$\mu_{{\rm LHY},\sigma}[n_{1,2}({\bf r},t)]=\delta E_{\rm LHY}/\delta n_\sigma$, where
\begin{equation}
E_{\rm LHY}=\frac{8}{15\sqrt{2\pi}}\Big(\frac{m}{4\pi\hbar^2}\Big)^{3/2}\int d\theta_k\sin\theta_k\sum_{\lambda=\pm}V_\lambda(\theta_k)^{5/2}\,
\end{equation}
is the LHY energy correction, with
\begin{equation}
V_\pm(\theta_k)=\sum_{\sigma=1,2} \eta_{\sigma\sigma}n_\sigma\pm \sqrt{(\eta_{11}n_1-\eta_{22}n_2)^2+4\eta_{12}^2n_1n_2}
\end{equation}
and $\eta_{\sigma\sigma'}=g_{\sigma\sigma'}+g^d_{\sigma\sigma'}(3\cos^2\theta_k-1)$, 
being $g^d_{\sigma\sigma'}=\mu_0\mu_{\sigma}\mu_{\sigma'}/3$ and $\theta_k$ 
the angle sustained by ${\bf k}$ with the $z$ axis. Since the dipole moments 
of the components are antiparallel, the inter-component dipolar potential is 
repulsive~(attractive) when the components are placed 
head-with-tail~(side-by-side)~(see Fig.~\ref{fig:1}(a)). As a result, the 
dipolar interaction strongly favors immiscibility, and a very large and 
negative $a_{12}$ is needed to drive the system miscible. In the following, 
we consider $a_{12}=150\,a_0$, but the actual value is irrelevant as long 
as the inter-component overlapping remains negligible.


\paragraph{Three-dimensionally self-bound ADM.--} 
We first consider the case of fully unconfined mixtures~($\omega_z=0$). 
As for parallel dipolar mixtures~\cite{Smith2021, Bisset2021}, an immiscible ADM 
may present a three-dimensionally self-bound solution, but of a 
markedly different nature. This is best understood in the impurity 
limit~($N_1\gg N_2$). Let us assume that component 1 forms a self-bound 
droplet with density $n_1({\bf r})$. The droplet exerts a potential 
$V^{1\to 2}_{\rm dd}({\bf r})=\int d^3r' V^{12}_{\rm dd}({\bf r}-{\bf r}') n_1({\bf r}')$ 
on component 2, which, as seen in Fig.~\ref{fig:1}(b), is characterized 
by a marked minimum at a given radius $\rho_0$, well outside the droplet. 
Particles in component 2 are trapped in this mexican-hat potential. 



\begin{figure}[t]
\centering
\includegraphics[width=\linewidth]{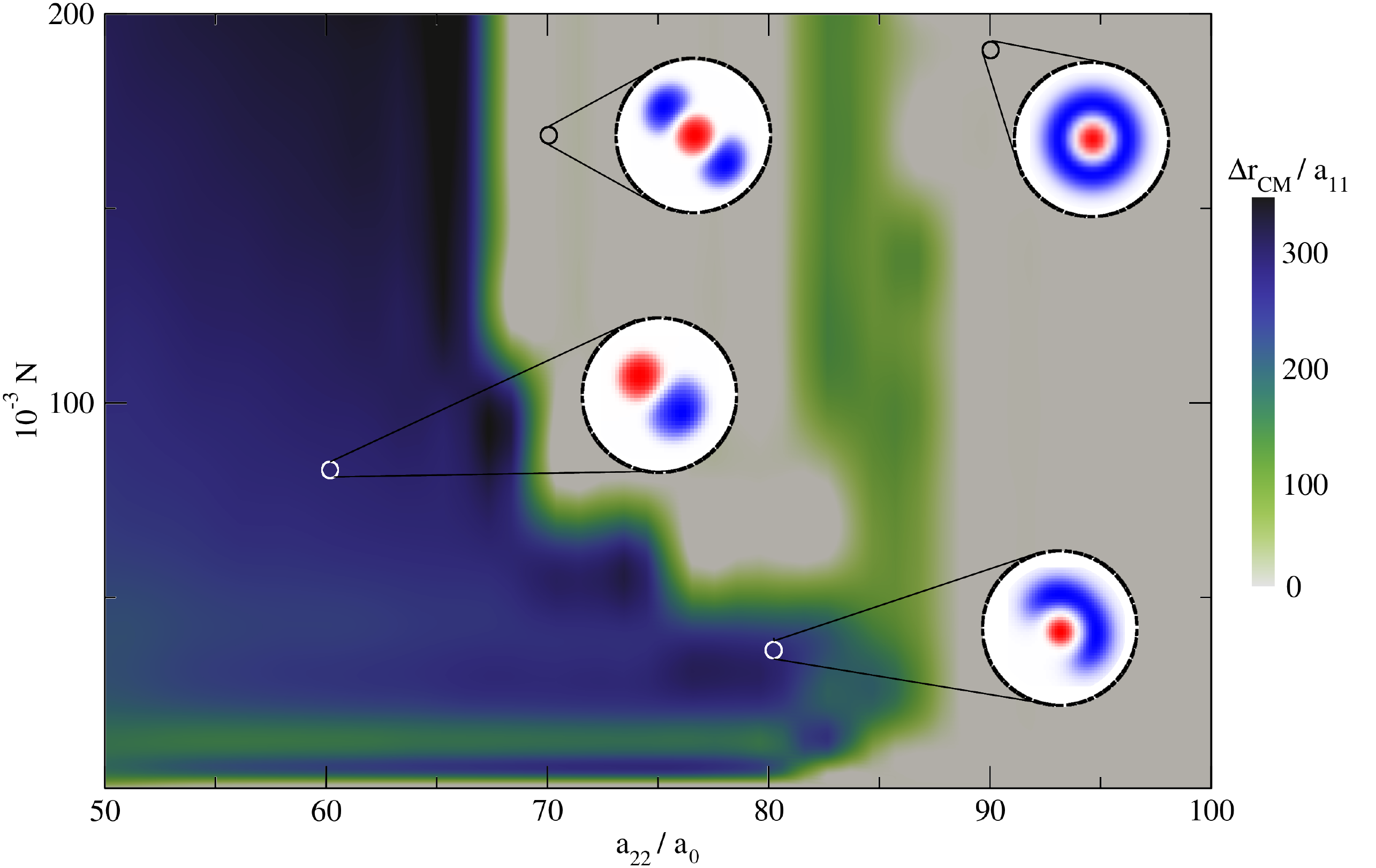}
\caption{Three-dimensionally self-bound ADMs. 
Ground-state configuration as a function of the total atom number $N$ 
and of $a_{22}$, for $a_{11}=50\,a_0$ and $N_{1,2}=N/2$. Whereas component 
$1$ always forms a single elongated droplet, component 2 may acquire 
different topologies, which we characterize using the separation 
$\Delta r_{\mathrm{CM}}$ between the center of masses of the two 
components~(color code). The different topologies are illustrated 
in the insets, where we depict the column density~(integrated over 
$z$) of the components, with red~(blue) indicating component 1~(2).}
\label{fig:2}
\end{figure}


In a more balanced mixture, the argument remains valid, but component 
$2$ also induces a similar potential $V^{2\to 1}_{\rm dd}({\bf r})$ on 
component $1$. Hence the two components confine each other mutually 
on the $xy$ plane, resulting in self-bound ADMs, as illustrated in 
Fig.~\ref{fig:2} for $N_{1,2}=N/2$, $a_{11}=50\,a_0$,  and different 
values of $a_{22}$ and $N$. For asymmetric intra-component interactions 
$a_{11}<a_{22}$, component $1$ remains a compact droplet, whereas the 
second component accommodates on the ring potential around the droplet. 
For low enough $a_{22}$, the energy is minimized by the formation of 
a single droplet in component 2, which for growing $N$ and $a_{22}$ 
spreads around the mexican-hat minimum until eventually forming a 
ring-like configuration. For intermediate $a_{22}$ values,  there 
is a second possible topology with two droplets of component 2 placed 
at opposite sides of the annular potential. 


\paragraph{Self-bound droplet crystals.--} 
When $\omega_z=0$, increasing the particle number $N$ results 
in more elongated solutions along the $z$ direction. As for 
single-component~(scalar) dipolar condensates~\cite{Boettcher2021}, 
this elongation is frustrated in the presence of a trap along 
$z$~($\omega_z>0$). In scalar condensates, this frustration 
results in the formation of multiple droplets. Although the 
droplets repel each other, the presence of a transversal trap 
on the $xy$ plane allows for the creation of 2D droplet 
crystals~\cite{Norcia2021, Bland2022a}. These crystals have 
however no intrinsic cohesion, and hence unravel in the absence 
of the $xy$ confinement.



\begin{figure}[t!]
\centering
\includegraphics[width=\linewidth]{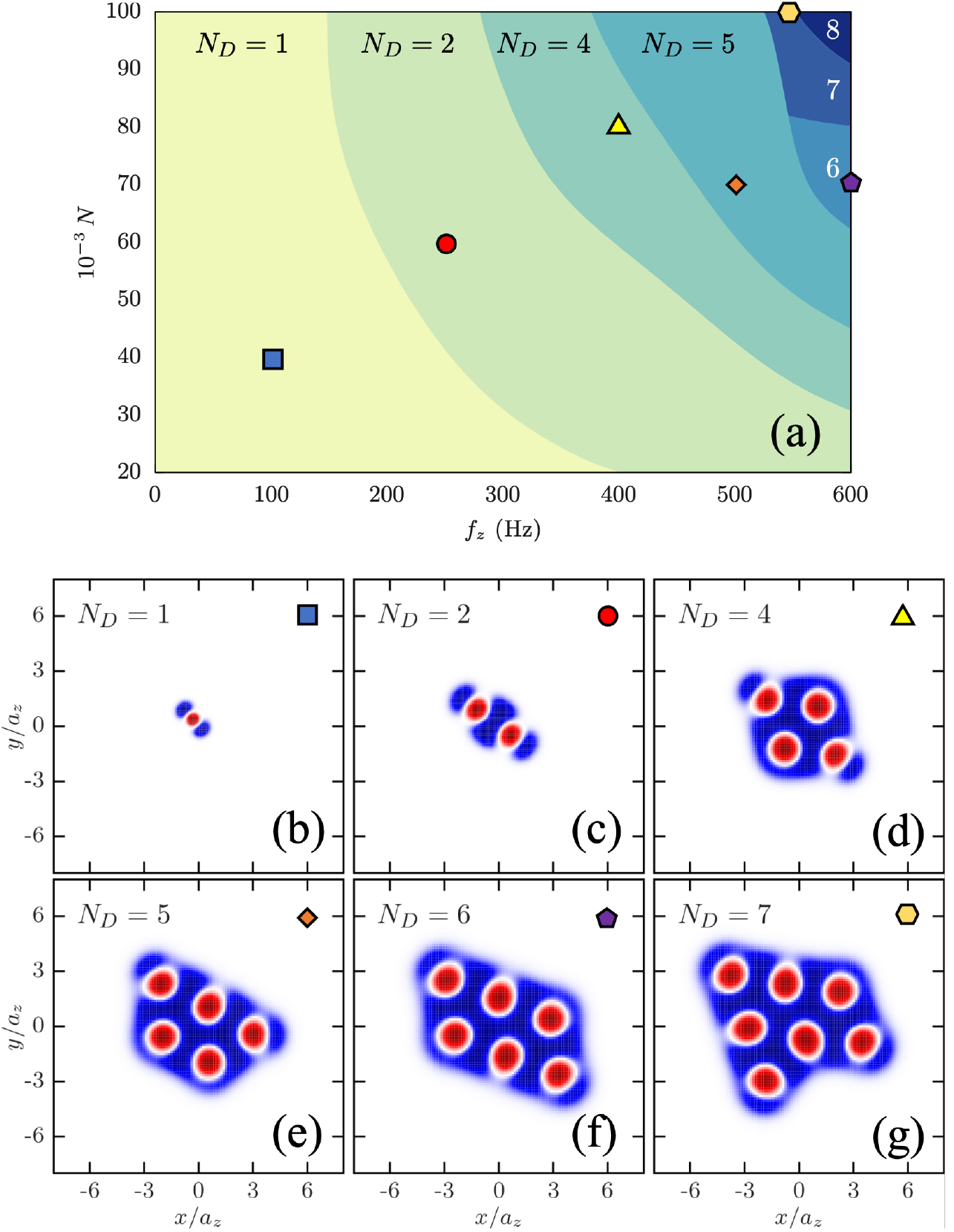}
\caption{Self-bound droplet crystals. (a) Phase diagram as a function 
of the atom number $N$ and the trap frequency $f_z=\omega_z/2\pi$, for 
$a_{11}=50\,a_0$ and $a_{22}=70\,a_0$. Colors correspond to configurations 
with a different number of droplets $N_D$ in component 1. Figures (b--g) 
show the column magnetization~(integrated along $z$) of the lowest-energy solution 
for selected cases, indicated with the corresponding symbol in Fig. (a). 
Red~(blue) regions are populated by  
component 1~(2).}
\label{fig:3}
\end{figure}



\begin{figure}[t!]
\centering
\includegraphics[width=\linewidth]{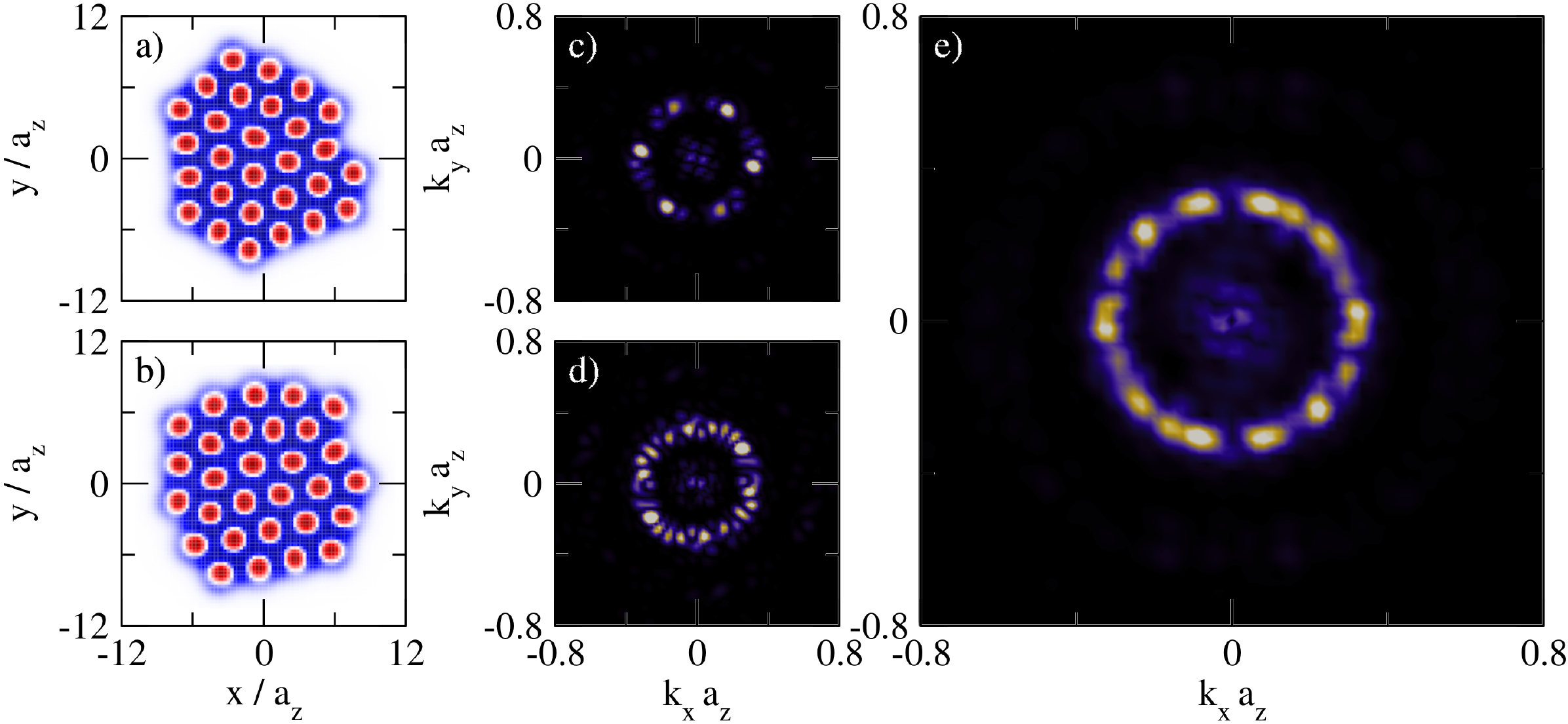}
\caption{(a,b) Single-shot realizations of the column magnetization for $a_{11}=50\,a_0$, 
$a_{22}=70\,a_0$, $N_{1,2}=5\times 10^4$ and $\omega_z/2\pi=1300$~Hz. Red~(blue) regions 
are populated by component 1~(2). (c,d) Corresponding momentum distribution for the second 
component, $\tilde n_2(k_x,k_y)$ in the $k_z=0$ plane, in the cases of Figs. (a) and (b), 
respectively. (e) Momentum distribution $\tilde n_2(k_x,k_y)$ averaged over $10$ different 
realizations.}
\label{fig:4}
\end{figure}


Remarkably, this is not the case in an ADM, as illustrated in 
Fig.~\ref{fig:3} for a balanced mixture $N_1=N_2$ and asymmetric 
intra-component interactions, $a_{11}=50\,a_0$ and $a_{22}=70\,a_0$. 
For a low-enough $\omega_z$, the three-dimensional solution~(with 
a single droplet in component 1) remains valid~(Fig.~\ref{fig:3}(b)). 
For an $N$-dependent critical $\omega_z$ the droplet splits into 
two. Each one of them exerts a mexican-hat potential on the second 
component, which gets trapped in the combined energy minimum. At 
the same time, crucially, the second component glues the two 
droplets together, forming a self-bound ADM~(Fig.~\ref{fig:3}(c)). 
As shown in Fig.~\ref{fig:3}(a), and illustrated for particular 
cases in Figs.~\ref{fig:3}(d--g), further increasing $\omega_z$ 
results in a growing number of droplets of component $1$ surrounded 
by a bath of component 2. In a scalar condensate, each droplet 
requires a minimal atom number to remain self-bound~(otherwise 
kinetic energy unbinds it), drastically limiting the total number 
of droplets. In contrast, in an ADM, droplets remain confined by 
the inter-component interaction, allowing for droplets with a much 
smaller number of atoms~\cite{Li2022, Bland2022b}. As a result, 
increasing $\omega_z$ results in 2D crystals with much more droplets 
compared to scalar condensates with the same total number of atoms.

We should emphasize that our results, based on imaginary-time evolution 
of Eq.~\eqref{eq:eGPE} with random initial conditions, reveal many 
possible solutions with very similar energy, which differ in the exact 
number and arrangement of the droplets~(see~\cite{SM}). We hence expect 
a significant experimental shot-to-shot variability, similar to that 
recently observed in experiments on 2D supersolids~\cite{Norcia2021}.

\paragraph{Interstitial superfluid.--} Due to the lack of overlapping, 
the droplets are mutually incoherent. In contrast, the component filling 
the crystal interstitials forms a superfluid that resembles, to some 
extent, the case of Helium in a porous medium~(although, in contrast to 
that scenario, droplets of component 1 do not form a rigid structure).
The approximately triangular crystalline structure of the droplets 
is inherited as well by the interstitial component 2, which builds 
hence a peculiar form of supersolid. The coherence and spatial density 
modulation of component 2 may be revealed in time-of-flight measurements. 
Figure~\ref{fig:4} shows the momentum distribution $\tilde n_2(k_x,k_y)$ 
in the $k_z=0$ plane. The approximate triangular structure~(Fig.~\ref{fig:4}(a)) results in an hexagonal pattern in the $\tilde n_2$ distribution~(Fig.~\ref{fig:4}(c)), although the above-mentioned variability of the exact droplet arrangement may result in a significant shot-dependent distortion~(see Figs.~\ref{fig:4}(b,d)). Note as well that, 
due to the lack of any confinement on the $xy$ plane, the patterns 
spontaneously break the polar symmetry and hence experience a random rotation 
from shot to shot. In any case, as expected from the theory of roton 
immiscibility~\cite{Wilson2012, Bland2022b}, the inter-droplet distance 
$R$ is fixed by the oscillator length $a_z=\sqrt{\hbar/m\omega_z}$. For 
the case of Fig.~\ref{fig:3}, $R\simeq 3\,a_z$ for all values of $N$ and 
$\omega_z$. This periodicity becomes evident from the average of the 
momentum distribution over many realizations, which shows a marked ring 
at $1/R$~(see Fig.~\ref{fig:4}(e)).



\begin{figure}[t!]
\centering
\includegraphics[width=\columnwidth]{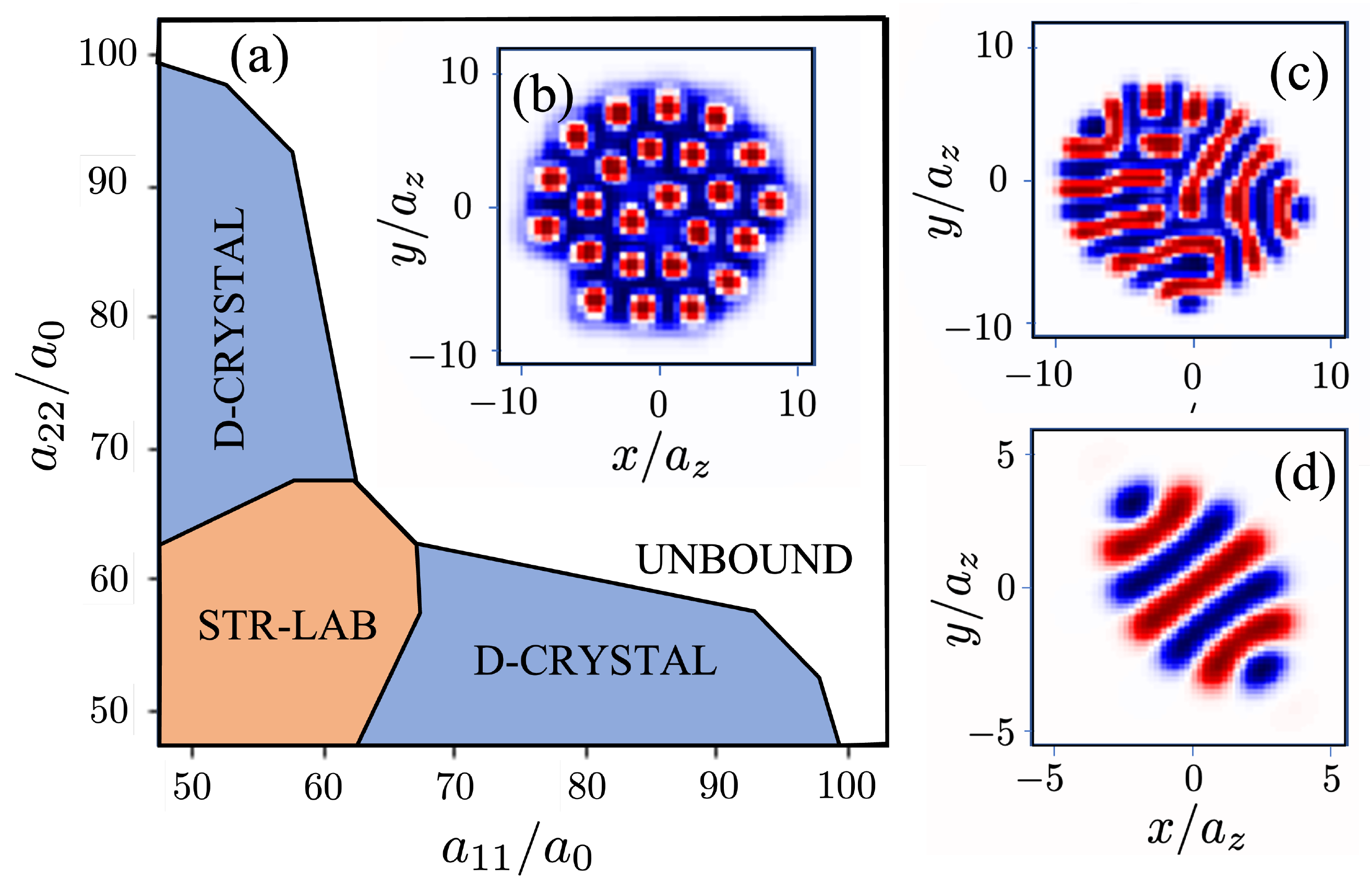}
\caption{(a) Phase diagram for $N_{1,2}=5\times 10^4$ and 
$\omega_z/2\pi = 1200$~Hz. Two different 
self-bound solutions are found: a droplet crystal (D-CRYSTAL), 
illustrated in Fig. (b) for $a_{11}=50\,a_0$ and $a_{22}=70\,a_0$, and 
a stripe/labyrinthic (STR-LAB) phase, illustrated in Fig. (c) 
for $a_{11}=55\,a_0$ and  $a_{22}=60\,a_0$. The case of a well-defined 
stripe phase is illustrated in Fig. (d), which has been evaluated 
for $\omega_z/2\pi=180$~Hz, $a_{11}=80\,a_0$, and $a_{22}=80\,a_0$. 
In Figs. (b--d) we depict the column magnetization. Red~(blue) regions 
are populated by component 1~(2).}
\label{fig:5}
\end{figure}


\paragraph{Crystal sublimation.--} For a fixed total number of 
particles, the cohesive energy decreases when the droplet number 
grows, since lowering the density reduces the inter-component dipolar 
attraction. Eventually, at a critical frequency $\omega_{z}^{\mathrm{cr}}$, 
the crystal unbinds, and both components evaporate. The critical 
frequency~($\omega_z^{\mathrm{cr}}/2\pi\simeq 1400$~Hz for the case 
on Fig.~\ref{fig:3}) is approximately determined as that for which 
the energy per particle reaches $\hbar\omega_z/2$, corresponding to 
an infinitely spread solution on the $xy$ plane. Interestingly, when 
$\omega_{z}$ approaches $\omega_{z}^{\mathrm{cr}}$, mutual attraction 
may still be enough to maintain a stable crystal, but insufficient 
to bind the whole interstitial superfluid, which hence partially 
evaporates~(see \cite{SM} for a more detailed discussion).

\paragraph{Self-bound stripe/labyrinthic patterns.--} 
Up to this point, we have considered a mixture with markedly asymmetric 
intra-component interactions. Interestingly, when $a_{11}\simeq a_{22}$, 
the mixture arranges in a different form of self-bound pattern~(note that 
$a_{11}=a_{22}$ if we consider a mixture of two maximally stretched 
magnetic states of the same atomic species). This is illustrated by the 
phase diagram of Fig.~\ref{fig:5}~(a), obtained for  $\omega_z/2\pi=1200$~Hz 
and $N_{1,2}=5\times 10^4$. For sufficiently large $|a_{11}-a_{22}|$, we 
obtain the above-mentioned droplet crystal~(Fig.~\ref{fig:5}(b)), which, 
as mentioned above, presents partial evaporation of the interstitial 
component in the vicinity of the unbinding threshold. In contrast, 
when $a_{11}\simeq a_{22}$ the mixture arranges in a labyrinthic 
phase, with a large shot-to-shot variability, formed by stripes 
with different orientations~(Fig.~\ref{fig:5}(c)). For lower trap 
frequencies, the ground-state configuration is given by a well-defined 
stripe crystal~(Fig.~\ref{fig:5}(d)). Note that in the labyrinthic/stripe 
phase both components form mutually incoherent domains.

\paragraph{Summary and Outlook.--}
Antiparallel dipolar mixtures allow for the formation of crystals 
with a genuine cohesive energy that remain self-bound in the absence 
of a transversal trap. The mutual confinement stems from the attractive 
inter-component interactions, and results in incoherent stripe/labyrinthic 
crystals in mixtures with symmetric intra-component interactions, and 
self-bound droplet crystals in asymmetric mixtures. The latter are 
particularly interesting, since while one component forms an approximately 
triangular array of incoherent droplets, the other component builds a 
superfluid in the interstitials, forming a peculiar form of supersolid 
that may be readily probed using time-of-flight measurements. Although 
we have considered the particular example of a dysprosium mixture, our 
results generally apply to other antiparallel magnetic or electric 
dipolar mixtures, including those of polar molecules.

The possibility of creating self-bound dipolar crystals opens 
intriguing perspectives for future studies, including the character 
of lattice excitations, which may remain self-bound or result in 
phonon evaporation~(resembling droplet evaporation in non-dipolar 
mixtures~\cite{Petrov2015}), the probing (e.g. by vortex formation) 
of the superfluidity of the interstitial component, as well as in 
general the exploration of the dynamics of self-bound crystals.


\acknowledgements
This work has been supported by Grant No. PID2020-114626GB-I00 (Ministerio de Ciencia 
e Innovaci\'on), by the European Union Regional Development Fund within 
the ERDF Operational Program of Catalunya (project QUASICAT/QuantumCat), 
by the Deutsche Forschungsgemeinschaft (DFG, German Research Foundation) 
under Germany’s Excellence Strategy–EXC-2123 QuantumFrontiers–390837967, 
and FOR 2247. M. A. is supported by FPI Grant PRE2018-084091. 


\bibliography{bibliography.bib}

\cleardoublepage
\appendix

\setcounter{equation}{0}
\setcounter{figure}{0}
\setcounter{table}{0}
\makeatletter
\renewcommand{\theequation}{S\arabic{equation}}
\renewcommand{\thefigure}{S\arabic{figure}}

%
\setcounter{equation}{0}
\setcounter{figure}{0}
\setcounter{table}{0}
\makeatletter
\renewcommand{\thefigure}{S\arabic{figure}}
\section{Supplementary Information}

The supplementary information contains additional details on the variability 
of the density patterns and the partial evaporation of the mixture mentioned 
in the main text.

\subsection{Shot-to-shot variability}
\begin{figure}[h!]
\centering
\includegraphics[width=\columnwidth]{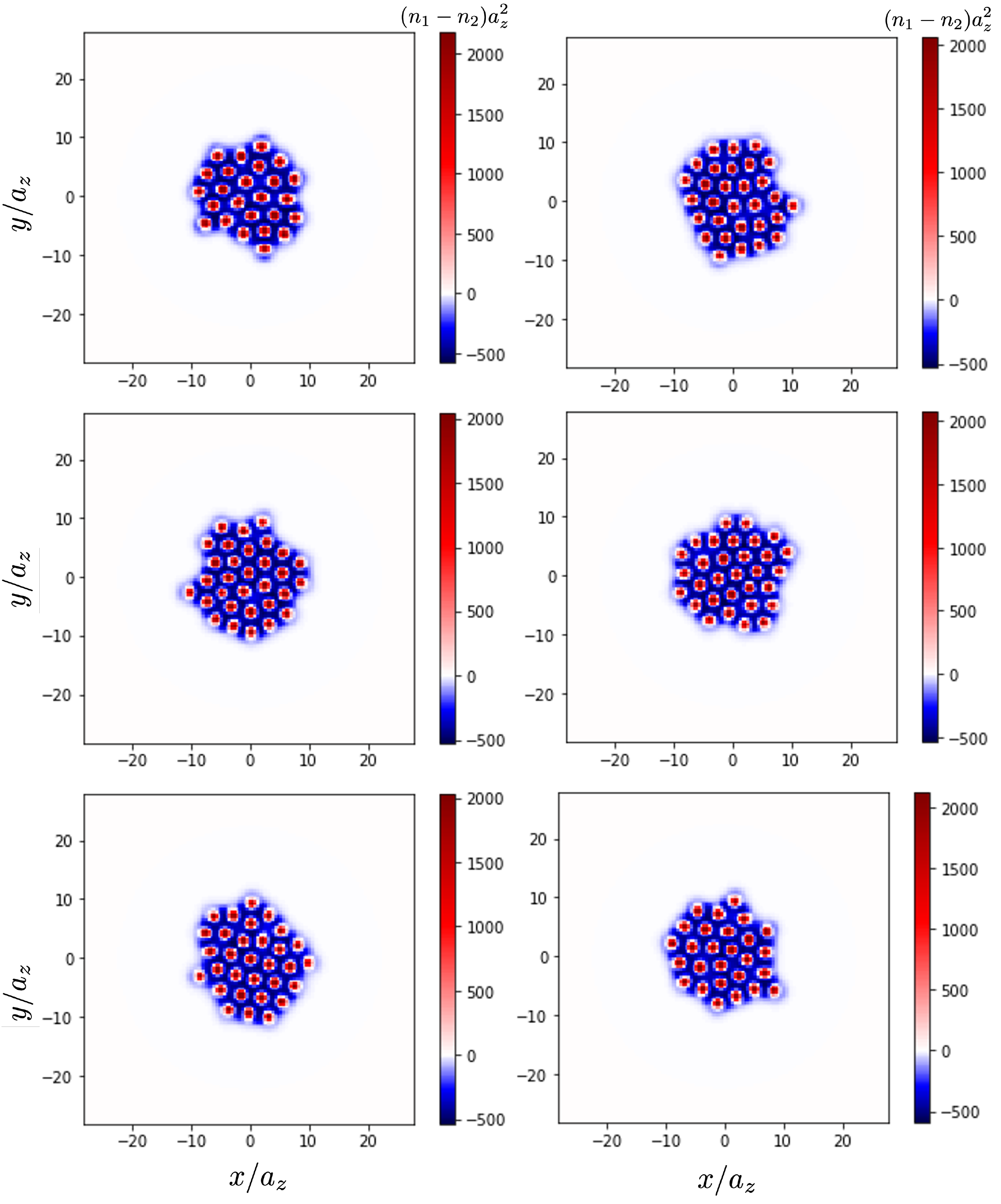}
\caption{Shot-to-shot variability. Different droplet lattice configurations 
obtained for the same parameters $a_{11}=50\,a_0$, $a_{22}=70\,a_0$, 
$\omega_z/2\pi=1200$~Hz, $N_{1,2}=5\times 10^4$. The plots show the 
column magnetization, with red~(blue) indicating component 1~(2).}
\label{fig:S1}
\end{figure}

There is a large shot-to-shot variability 
of the exact number of droplets and their arrangement in the droplet crystal. 
We illustrate this point with Fig.~\ref{fig:S1}, where we show different 
configurations for the same parameters $a_{11}=50\,a_0$, $a_{22}=70\,a_0$, 
$\omega_z/2\pi=1200$~Hz, and $N_{1,2}=5\times 10^4$. The configurations, 
which have an energy per particle $E/N\simeq 0.22\,\hbar\omega_z$, differ in energy by 
less than $1\%$, and have a number of droplets ranging from $23$ to $31$. 



Note that the droplet arrangement is only approximately triangular, 
and the actual distribution, which is generally non-uniform, may 
significantly depart from a triangular lattice. This is reflected 
in the structure of the interstitial component, which results in 
the distorted momentum distributions depicted in Fig. 4 of the main 
text. A similar shot-to-shot variability is observed in the 
stripe/labyrinthic phase.

\subsection{Evaporation of the droplet crystal}

\begin{figure}[h!]
\centering
\includegraphics[width=\linewidth]{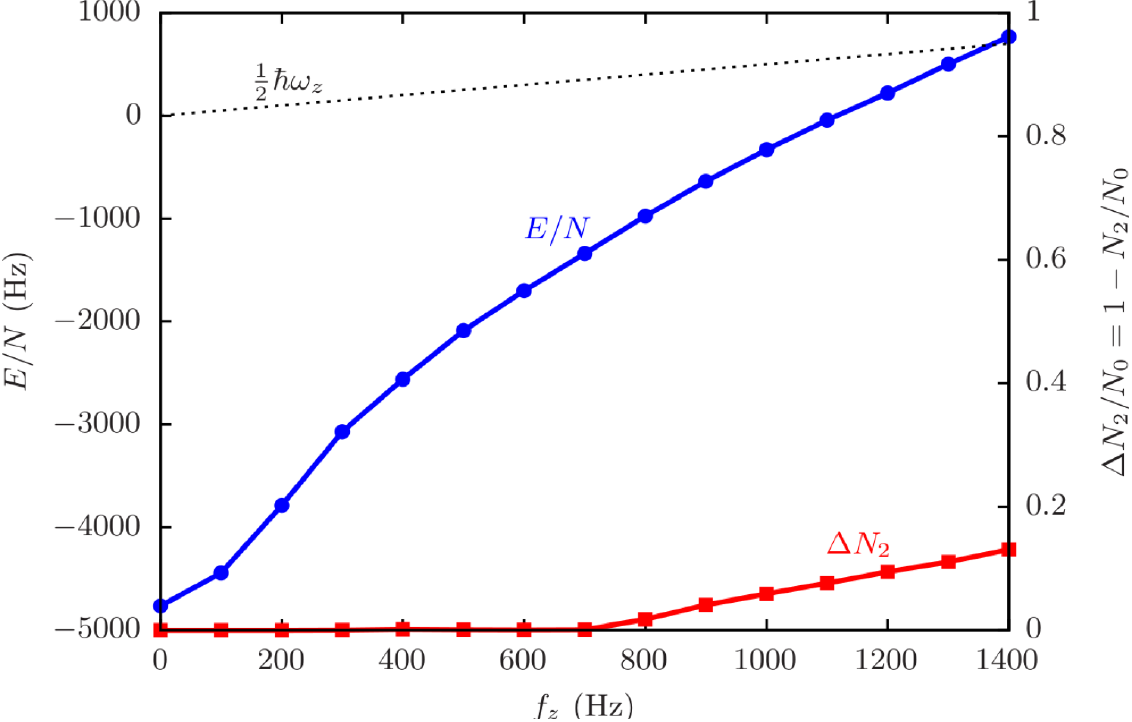}
\caption{Energy per particle~(blue circles) as a function of the trap 
frequency for $a_{11}=50\,a_0$, $a_{22}=70\,a_0$ and $N_{1,2}=5\times 10^4$. 
The dashed line depicts the energy per particle~($\hbar\omega_z/2$) 
corresponding to an infinitely spread mixture on the $xy$ plane. The 
red squares indicate the proportion of atoms in component 2 which are 
evaporated~(see text).}
\label{fig:S2}
\end{figure}

When $\omega_z$ increases, 
the number of droplets grows and the density decreases. As 
a result, the cohesive energy is reduced, and the self-bound 
solution eventually unbinds. Figure~\ref{fig:S2} shows the 
energy per particle as a function of the trap frequency 
for $a_{11}=50\,a_0$, $a_{22}=70\,a_0$ and $N_{1,2}=5\times 10^4$. 
The unbinding of the droplet crystal occurs approximately 
when the energy per particle $E/N$ reaches $\hbar\omega_z/2$, 
which, for the case of Fig.~\ref{fig:S2}, occurs at 
$\omega_z^{\mathrm{cr}}/2\pi\simeq 1400$~Hz. Indeed, beyond 
that value we do not find well defined self-bound solutions 
in our simulations.

When $\omega_z$ approaches $\omega_z^{\mathrm{cr}}$, the crystal 
remains bound, but the interstitial component may present partial evaporation. 
In order to take this into account we considered absorbing boundary 
conditions in our imaginary-time simulations. We fixed a given radius 
$\rho_{c}$ on  the $xy$ plane, such that the crystal is well contained 
in a circle of radius $\rho<\rho_{c}$. Particles that reach $\rho>\rho_c$ 
during imaginary-time evolution are considered as evaporated. We 
indicate in Fig.~\ref{fig:S2} the proportion of the interstitial 
component that is evaporated, which as expected grows when approaching 
$\omega_z^{\mathrm{cr}}$. In the unbound regime, the whole mixture 
eventually leaves (in imaginary time) the region $\rho<\rho_c$.



\end{document}